\documentclass{moriond}

\usepackage{graphicx}
\usepackage{float}
\usepackage{amsmath}

\def\be{\begin{equation}}
\def\ee{\end{equation}}
\def\bea{\begin{eqnarray}}
\def\eea{\end{eqnarray}}

\begin{document}
\vspace*{4cm}
\title{Global analysis of the $U(3)^5$ symmetric SMEFT}
\author{ Riccardo Bartocci }
\address{PRISMA+ Cluster of Excellence \& Mainz Institute for Theoretical Physics\\ Johannes Gutenberg University, D-55099 Mainz, Germany}

\maketitle\abstracts{The $U(3)^5$ symmetry within the SMEFT framework restricts the inclusion of only fully flavor-conserving operators at dimension six. This proceeding presents a global analysis of the SMEFT under this assumption. We provide global constraints on all 41 Wilson coefficients, utilizing leading-order and next-to-leading-order SMEFT predictions for various experiments including parity-violating experiments, Electroweak Precision Observables (EWPO), Higgs physics, top quark interactions, flavor observables, dijet production, and lepton scatterings. We address issues concerning the constraints of specific four-quark operators, investigate correlations between observables at different energy scales, and assess the impact of next-to-leading-order contributions on the global fit.}

\section{Introduction}

While the Standard Model of particle physics has proven effective in describing particle interactions observed at colliders, it falls short in explaining various phenomena. Therefore, in the absence of direct evidence of New Physics (NP), the Standard Model Effective Field Theory (SMEFT) emerges as a valuable tool for rigorously parameterizing the effects of NP within the energy scale accessible by current experiments.
Within SMEFT, the dimension-four Standard Model Lagrangian is extended including higher-dimensional operators involving only Standard Model fields and preserving the symmetries of the Standard Model. Conducting global fits of all dimension-six SMEFT operators is impractical due to the big number of independent Wilson coefficients. However, exploiting flavour symmetry allows for the reduction of independent parameters at the high scale.
Given that flavor-violating observables already impose significant constraints on flavor-violating coefficients, it is reasonable to mitigate them through the assumption of minimal flavor violation (MFV), where only the Yukawa couplings serve as sources of $U(3)^5$ breaking.
In this study, we consider an exact $U(3)^5$ flavor symmetry for dimension-six operators at high scales. With this assumption, combined with CP-symmetry, the number of independent operators is reduced from 2499 to 41.
Our analysis utilizes data from various experiments including electroweak precision observables (EWPO), low-energy parity violation experiments, Higgs physics, top quark interactions, flavor observables, Drell-Yan (DY), and dijet production. Through these datasets, we perform a working global fit without any remaining flat directions. Since certain coefficients are poorly constrained at leading order (LO), we also incorporate next-to-leading order (NLO) contributions to the observables, highlighting the impact of these loop corrections on the fit.

\section{SMEFT and flavour symmetry}
\label{sec:symmetricSMEFT}

The SMEFT Lagrangian, at order $1/\Lambda^2$ is expressed as
\begin{equation}
\mathcal{L}_{\text{SMEFT}}=\mathcal{L}_{\text{SM}}+\sum_i \frac{C_i}{\Lambda^2} \, Q_i,
\label{eq:SMEFT}
\end{equation}
where $C_i$ are the Wilson coefficients of the dimension six operators $Q_i$ and $\Lambda=4$ TeV (in this work) denotes the heavy scale associated with NP. The SMEFT theory predictions are considered only at linear order in the Wilson coefficients. Quadratic SMEFT contributions are suppressed by $1/\Lambda^{4}$ and therefore as power suppressed as the dimension-eight linear contributions. 
The SMEFT serves as a tool for model independent studies of NP and a priori all the flavour structures are allowed. However, beyond the SM flavour violation is already severely constrained, therefore symmetries for the flavour sector can be assumed to reduce the amount of BSM flavour violation.
We consider a complete flavour symmetry, namely $U(3)^5$, of the dimension six SMEFT operators, given by
\begin{equation}
U(3)^5=U(3)_\ell \times U(3)_q\times U(3)_e\times U(3)_u\times U(3)_d,
\end{equation}
where $\{\ell, q, e, u, d\}$ are the SM fermions.
This assumption is called the \textit{minimal} MFV because it contains the minimum amount of $U(3)^5$ breaking which comes from the Yukawa couplings of the dimension four Lagrangian.
Once this assumption is imposed, there are 41 independent CP-conserving SMEFT coefficients~\cite{Faroughy:2020ina}. 
Even if the dimension six operators are perfectly flavour symmetric, flavour violating observables are still generated via the renormalisation group flow: since the Yukawa matrices break the symmetry, the flavour violating dimension six operators can be produced at low scales, but their coefficients will depend only on flavour conserving high scale Wilson coefficients.

\section{Leading order datasets and dijets production}
\label{sec:datasets}

\begin{figure}[H]
    \centering   \includegraphics[width=0.8\textwidth]{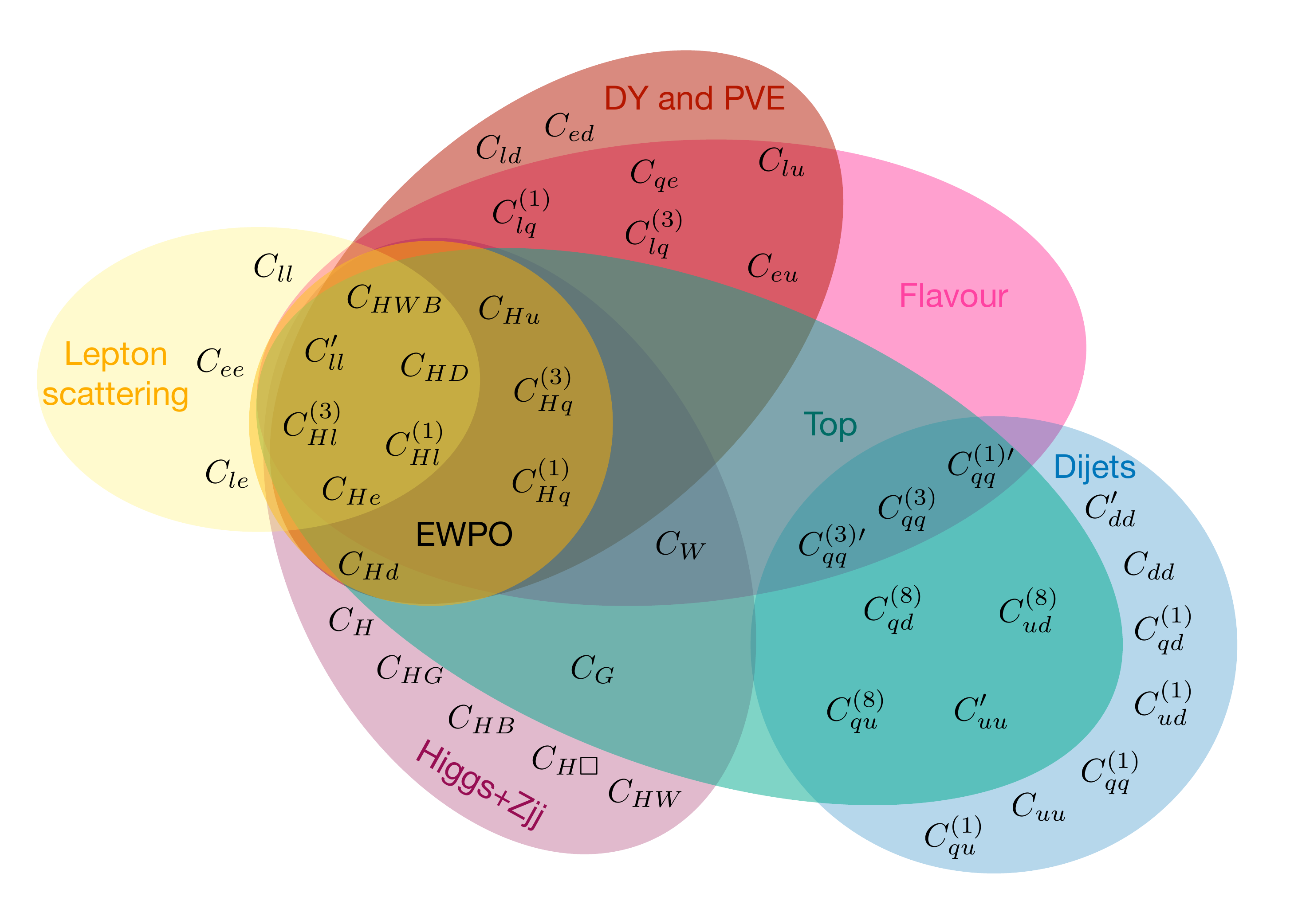}
    \caption{Sets of operators contributing to the different datasets at leading order.}
    \label{fig:venn_diag}
\end{figure}
The global fit includes the following datasets: EWPO~\cite{Dawson:2022bxd}, Higgs as in ~\cite{Anisha:2021hgc}, top~\cite{Ellis:2020unq}, low-energy parity violation experiments (PVE) and lepton scattering~\cite{Falkowski:2017pss}, flavour observables~\cite{Straub:2018kue}, Drell-Yan production~\cite{Allwicher:2022mcg} and dijet+photon production~\cite{Bartocci:2023nvp}. 
How the different operators appear in the various datasets can be seen in Figure~\ref{fig:venn_diag}. 

\subsection{Theory predictions for dijet+photon production}
Some four-quark operators are particularly hard to constrain: top and flavour cannot set bounds on them. Therefore, an additional observable is needed and a natural candidate is dijets production at the LHC.
However, due to the triggers for jets at the LHC, only very high energy data are available, in particular above the multi-TeV invariant masses. 
At so high energies, the terms quadratic in the SMEFT coefficients become bigger than the linear ones and this brings inconsistencies in the treatment of these quadratic contributions as theoretical uncertainty.
To overcome the issue, we considered the production of two jets associated with a photon~\cite{ATLAS:2019itm}. This slightly different process enables us to have acces to lower dijet invariant-mass, in particular $m_{jj}< 1.1$~TeV.
Through Madgraph simulations and SMEFTsim \cite{Brivio:2020onw}, we have evaluated the dimension six linear and quadratic SMEFT predictions for the differential cross section of this process, showing that the quadratic contributions are kept under control in this energy range.

\section{Global analysis results and comparison between LO and NLO}

\begin{figure}
    \centering
    \includegraphics[width=0.99\textwidth]{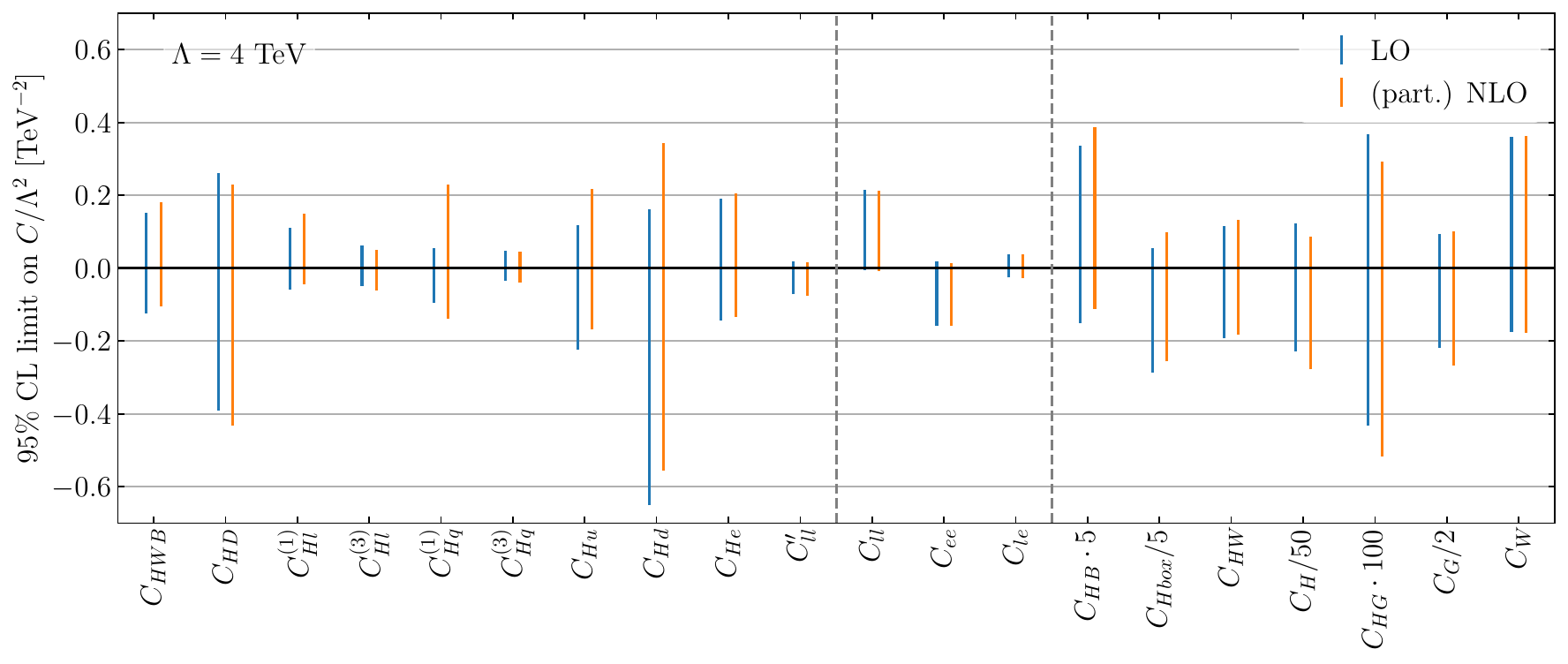}
    \includegraphics[width=0.95\textwidth]{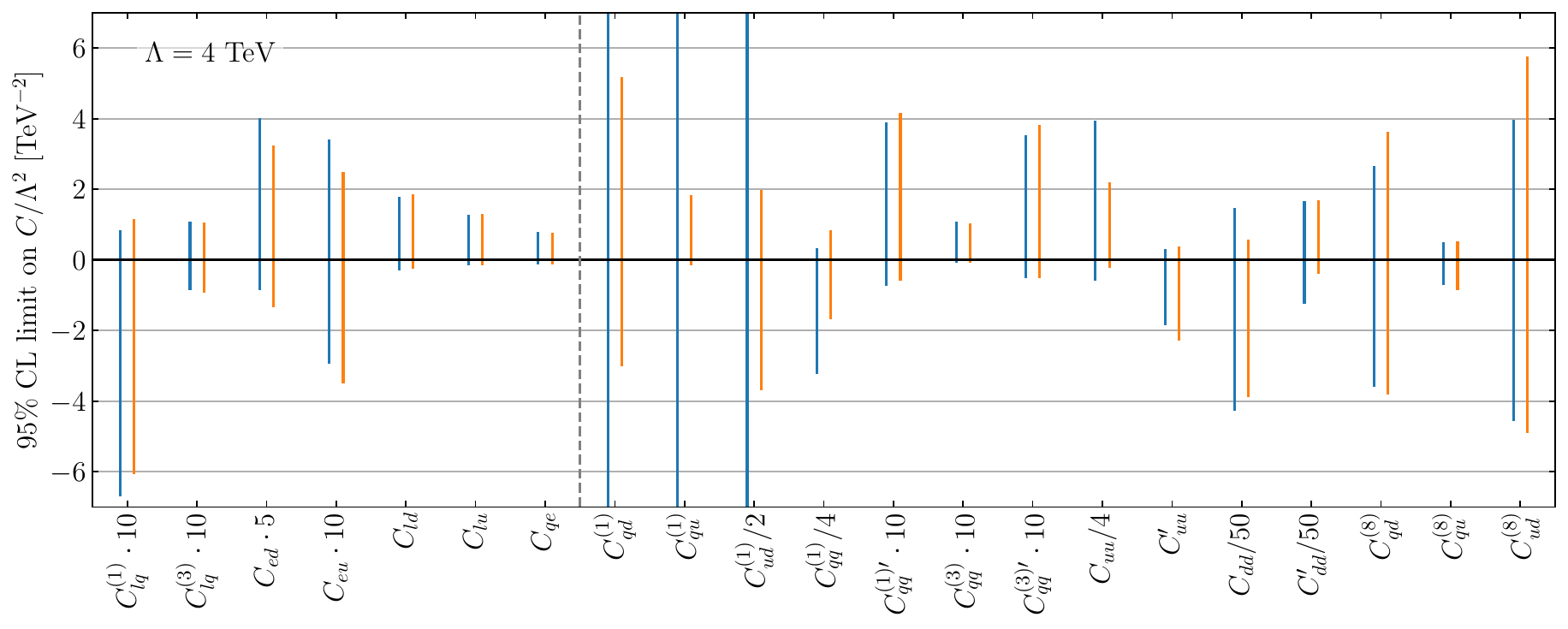}
    \caption{Comparison of the LO and NLO global fit}
    \label{fig:LO_vs_NLO_fit}
\end{figure}

In Figure~\ref{fig:LO_vs_NLO_fit}, we present the results of both the leading order (LO) and next-to-leading order (NLO) fits. Upon considering NLO contributions, significant improvements are observed in the bounds for $C_{qd}^{(1)}$, $C_{qu}^{(1)}$, and $C_{ud}^{(1)}$. Initially, these three operators are constrained solely by dijet data at LO. However, with the inclusion of NLO effects, Higgs, Top, and EWPO datasets also contribute to setting bounds on these coefficients. Consequently, we observe an improvement of approximately two orders of magnitude compared to the LO fit. Despite the expectation of increased correlations among Wilson coefficients at NLO, leading to potential impacts on previously well-constrained operators at LO, we find that the bounds on other coefficients remain relatively stable in the NLO fit. Notably, the bounds on the ten LO EWPO operators exhibit remarkable stability.

In the NLO fit, all bounds lie below $|C|/\Lambda^2 < 10/\text{TeV}^2$ with $95\%$ confidence level, with the exception of the four-quark operators $C_{dd}$ and $C_{dd}^\prime$. Among the EW operators, the only one notably affected is $C_{Hq}^{(1)}$, whose bound weakens by a factor of 2 due to correlations with certain four-quark operators, notably with $C_{qq}^{(1)}$ and $C_{uu}$, as detailed in \cite{Bartocci:2023nvp}.
In conclusion, all 41 Wilson coefficients of the $U(3)^5$ symmetric SMEFT are compatible with the SM within~$2\sigma$ in our global fit.

\section*{Acknowledgments}

I would like to express my gratitude to my collaborators, Anke Biekötter and Tobias Hurth, for their fruitful teamwork on this project. Many thanks also go to the organizers of Moriond 2024 EW for their excellent management of the conference. This work is supported by  the  Cluster  of  Excellence  ``Precision  Physics,  Fundamental
Interactions, and Structure of Matter" (PRISMA$^+$ EXC 2118/1) funded by the German Research Foundation (DFG) within the German Excellence Strategy.

\section*{References}
\bibliography{bibliography}

\end{document}